\documentclass[iop,revtex4]{emulateapj}
\usepackage{pdflscape}
\usepackage{longtable}

\def\spose#1{\hbox to 0pt{#1\hss}}
\def\lta{\mathrel{\spose{\lower 3pt\hbox{$\mathchar"218$}}
     \raise 2.0pt\hbox{$\mathchar"13C$}}}
\def\gta{\mathrel{\spose{\lower 3pt\hbox{$\mathchar"218$}}
    \raise 2.0pt\hbox{$\mathchar"13E$}}}

\shorttitle{Chemical Signatures of the First SNe in Sculptor}
\shortauthors{Simon et al.}
\slugcomment{Accepted for publication in ApJ}

\begin{document}

\title{Chemical Signatures of the First Supernovae in the
  Sculptor Dwarf Spheroidal Galaxy\altaffilmark{*}}

\author{Joshua D. Simon\altaffilmark{1}, 
  Heather R. Jacobson\altaffilmark{2},
  Anna Frebel\altaffilmark{2},
  Ian B. Thompson\altaffilmark{1}, 
  Joshua J. Adams\altaffilmark{3}, and
  Stephen A. Shectman\altaffilmark{1}
}

\altaffiltext{*}{This paper includes data gathered with the 6.5 meter
  Magellan Telescopes located at Las Campanas Observatory, Chile.}

\altaffiltext{1}{Observatories of the Carnegie Institution of
  Washington, 813 Santa Barbara St., Pasadena, CA 91101;
  jsimon,ian,jjadams,shec@obs.carnegiescience.edu}

\altaffiltext{2}{Kavli Institute for Astrophysics and Space Research
  and Department of Physics, Massachusetts Institute of Technology, 77
  Massachusetts Avenue, Cambridge, MA 02139; hjr,afrebel@mit.edu}

\altaffiltext{3}{ASML US, 77 Danbury Rd., Wilton, CT 06897};

\begin{abstract}
We present a homogeneous chemical abundance analysis of five of the
most metal-poor stars in the Sculptor dwarf spheroidal galaxy.  We
analyze new and archival high resolution spectroscopy from
Magellan/MIKE and VLT/UVES and determine stellar parameters and
abundances in a consistent way for each star.  Two of the stars in our
sample, at [Fe/H]$ = -3.5$ and [Fe/H]$ = -3.8$, are new discoveries
from our Ca~K survey of Sculptor, while the other three were known in
the literature.  We confirm that Scl~07-50 is the lowest metallicity
star identified in an external galaxy, at [Fe/H]$ = -4.1$.  The two
most metal-poor stars both have very unusual abundance patterns, with
striking deficiencies of the $\alpha$ elements, while the other three
stars resemble typical extremely metal-poor Milky Way halo stars.  We
show that the star-to-star scatter for several elements in Sculptor is
larger than that for halo stars in the same metallicity range.  This
scatter and the uncommon abundance patterns of the lowest metallicity
stars indicate that the oldest surviving Sculptor stars were enriched
by a small number of earlier supernovae, perhaps weighted toward
high-mass progenitors from the first generation of stars the galaxy
formed.
\end{abstract}


\section{INTRODUCTION}
\label{intro}

Metal-poor stars represent the local equivalent of the high-redshift
universe and supply us with a uniquely detailed view of the conditions
in early galaxies.  The lowest metallicity stars in dwarf galaxies are
particularly enticing targets for chemical abundance studies because
dwarfs have simpler merging and evolutionary histories than more
massive systems like the Milky Way.

The first extremely metal-poor (EMP) stars with $\rm{[Fe/H]} \le -3.0$
in nearby dwarf galaxies were discovered by \citet{kirby08}.  Over the
past five years, a sizable sample of such stars has been identified
and analyzed, in both the ultra-faint dwarfs and the brighter
classical dwarf spheroidals
\citep*{aoki09,ch09,ch10,fks10,frebel10,simon10,norris10a,norris10b,tafelmeyer10,kc12,gilmore13,starkenburg13,fsk14,ishigaki14}.
The initial results of these studies were that EMP stars in dwarf
galaxies appear remarkably similar to EMP stars in the Milky Way,
suggesting that early chemical evolution is largely independent of
galactic environment \citep[e.g.,][]{fks10,frebel10,simon10}.  Now
that a statistically significant number of EMP stars beyond the Milky
Way are available, we can begin to examine this conclusion more
carefully, and search for outliers from the typical abundance pattern
rather than simply characterizing broad trends within the population.

In this paper we present a homogeneous detailed chemical abundance
analysis of five of the most metal-poor stars in the Sculptor dwarf
spheroidal (dSph), with $\rm{[Fe/H]} \le -3.2$.  Three of these stars
have been studied previously by \citet[][hereafter T10]{tafelmeyer10}
and \citet{fks10}, but we re-analyze those data with a common set of
methods and assumptions.  We also add two new EMP stars identified on
the basis of their weak Ca~K absorption in our Magellan search for the
most metal-poor stars in the Milky Way's southern dSphs, and determine
their chemical abundance patterns with high resolution spectroscopy
for the first time.  Additional descriptions of the survey and the
resulting sample of EMP stars will be provided in future papers.
In Section~\ref{observations} we describe the data used in this study
and the reduction methods.  We present the chemical abundance
measurements in Section~\ref{results}, and we discuss the abundance
patterns and their implications for the early history of Sculptor in
Section~\ref{discussion}.  We summarize the paper in
Section~\ref{conclusions}.

\section{OBSERVATIONS AND DATA REDUCTION}
\label{observations}

We selected Scl~6\_6\_402 and Scl~11\_1\_4296 as candidate Sculptor
members from the photometry catalog of \citet*{coleman05}.  We
obtained $R \approx 700$ spectra centered on the Ca~K line of
$\sim2000$ such stars with the f/2 camera of the IMACS spectrograph
\citep{imacs} from $2009-2011$.  These were two of the stars with the
smallest Ca~K equivalent widths in the sample, marking them as likely
EMP stars.  We obtained medium resolution Magellan/MagE \citep{mage}
spectra of them in 2010 December, confirming their extremely low
metallicity.  We then observed the two stars with Magellan/MIKE
\citep{mike} at $R \approx 25000$ over a wavelength range of
$3460-9410$~\AA on ten nights between 2013 October and 2014 June,
accumulating a total of 20~hr of integration time on Scl~6\_6\_402 and
9.7~hr on Scl~11\_1\_4296.  The MIKE spectra were reduced with the
Carnegie Python routines originally described by \citet{kelson03}.

Two other EMP stars in Sculptor, Scl~07-49 and Scl~07-50, were
identified by T10 in the data set of the Dwarf Abundances and Radial
velocities Team \citep{tolstoy04,helmi06} using the Ca triplet
metallicity calibration from \citet{starkenburg10}.  T10 obtained $R
\approx 40000$ spectra of these stars with VLT/UVES \citep{uves} and
measured their chemical abundances.  The spectrum of Scl~07-50 was
obtained with both the red and blue arms of UVES, covering
$3700-10200$~\AA, while Scl~07-49 was observed only with the red arm
($4700-10200$~\AA).  To ensure that the full sample can be placed on a
common abundance scale we downloaded the reduced spectra from the ESO
archive and analyzed them with the same method we followed for the
other stars (see \S~\ref{results}).  The data available online are
individual exposures processed by version 5.1.5 of the UVES pipeline,
which handles standard reduction procedures and merges the echelle
orders.  We then normalized the spectra and coadded the frames for
each star.  For further details on the observations, see T10.

S1020549 was identified by \citet{kirby09} as an EMP star at [Fe/H]$ =
-3.8$ (at the time the most metal-poor star in an external galaxy)
based on medium resolution Keck/DEIMOS spectroscopy.  \citet{fks10}
confirmed that result and analyzed its chemical abundance pattern with
an $R \approx 33000$ Magellan/MIKE spectrum covering the same
wavelength range as our new MIKE data.  We use that spectrum again
here, but as for Scl~07-49 and Scl~07-50 we re-measure the equivalent
widths and re-determine the stellar parameters and chemical
abundances.

\section{DETERMINATION OF STELLAR PARAMETERS AND CHEMICAL ABUNDANCES}
\label{results}

\subsection{Measurement Procedures}

We measured the equivalent widths (EWs) of metal lines in the spectra
by fitting a Gaussian to each absorption line and integrating the area
under the Gaussian.  The lines were selected from the line list
constructed by \citet{roederer08}.  Where available, we have adopted
collisional damping constants from \citet{barklem00}; for other lines
we used the \citet{unsold55} approximation.  We set the continuum of
each spectrum by fitting a low-order cubic spline to each spectral
order.  Because the spectra of these faint stars have low S/N, the
placement of the continuum can be challenging, especially at blue
wavelengths.

Our derivation of the stellar parameters closely follows that of
\citet{fsk14} and \citet{frebel13}: using the analysis code of
\citet{casey14}, we employ $\alpha$-enhanced one-dimensional
plane-parallel ATLAS9 model atmospheres from \citet{ck04} and the MOOG
stellar analysis code \citep{sneden73} updated to include a treatment
of Rayleigh scattering \citep{sobeck11}.  We assume local
thermodynamic equilibrium throughout this paper, except where
otherwise noted.  Starting from the EW measurements, we computed
effective temperatures and surface gravities in the usual manner by
enforcing ionization and excitation balance of iron line abundances.
The microturbulent velocity was derived iteratively by minimizing the
trend of \ion{Fe}{1} abundance with reduced equivalent width.  We then
corrected these spectroscopic parameters according to the prescription
of \citet{frebel13} to place the measurements on the same scale as
studies that calculate stellar parameters from photometry alone.  As a
check of this process, we also calculated photometric temperatures
using $V - K_{s}$ colors determined from the optical photometry of
\citet{coleman05} and the near-infrared photometry of
\citet{menzies11} and the \citet*{alonso99} color-temperature
relation.  The photometric temperatures of Scl~11\_1\_4296,
Scl~6\_6\_402, and S1020549 are within 100~K of our adopted values.
For Scl~07-50 and Scl~07-49 the colors suggest temperatures that are
$200-300$~K warmer.  T10 find a temperature for Scl~07-49 that is in
agreement with ours rather than the photometric value, but they prefer
a higher temperature for Scl~07-50.\footnote{The $K_{{\rm s}}$
  magnitudes for these two stars are $\sim0.16$~mag fainter in
  \citet{menzies11} than in the VISTA commissioning data reported by
  T10, which could account for a temperature shift of $\sim200$~K, but
  neither paper contains enough information about the photometric
  calibration to determine which one is correct.}  Because the quality
of the spectrum for Scl~07-50 is relatively high and so many iron
lines can be included in our analysis, we consider the
spectroscopically derived temperature more likely to be accurate.  We
also note that \citet{frebel13} showed that the microturbulent
velocity determined with this method is in good agreement with values
obtained by other authors in the literature.

We list the derived stellar parameters for each star in
Table~\ref{sparams}, the EWs in Table~\ref{ew_table}, and the
abundances in Table~\ref{abundance_table}.  All abundance ratios have
been calculated using the \citet{asplund09} solar abundance scale.  We
estimate the systematic uncertainties on the derived abundances by
adjusting the atmospheric parameters by their uncertainties (see
Table~\ref{sparams}) and re-determining the abundance ratios.  The
uncertainties on each element as a result of changing $\rm T_{eff}$,
$\log{g}$ and the microturbulent velocity for each star are listed in
Table~\ref{uncert}.  The statistical uncertainties for elements with
multiple lines are defined to be equal to the dispersion in the
abundance ratio about the mean value divided by the square root of the
number of lines.  In cases where this dispersion is unrealistically
small, we impose a minimum value of 0.10~dex.  Following
\citet{frebel10}, who analyzed similar stars with spectra of similar
quality, we adopt a minimum abundance uncertainty of 0.2~dex for
elements where abundances are measured from a single line.  The total
uncertainty for each element is the sum in quadrature of the above
terms.

\begin{deluxetable*}{lcccccccccc} 
\tablecolumns{11} 
\tablewidth{0pt} 
\tabletypesize{\small}
\tablecaption{Coordinates and Stellar Parameters}
\tablehead{
\colhead{Star}  &
\colhead{R.A. (J2000)}  &
\colhead{Decl. (J2000)}  &
\colhead{$V$}  &
\colhead{$V-I$}  &
\colhead{$V-K_{{\rm s}}$}  &
\colhead{T$_{\rm{eff}}$ } & 
\colhead{$\log{g}$ }    & 
\colhead{$v_{\rm{micr}}$ }  & 
\colhead{$\mbox{[Fe/H]}$ } &
\colhead{S/N\tablenotemark{a}}\\
\colhead{} &
\colhead{} &
\colhead{} &
\colhead{} &
\colhead{} &
\colhead{} &
\colhead{[K]} &
\colhead{[dex]} & \colhead{[km~s$^{-1}$]} & \colhead{[dex]}  &
\colhead{} }
\startdata
Scl~11\_1\_4296 & 00:59:38.75 & $-33$:46:14.6 & 19.16 & 1.12 & 2.31 & $4770 \pm 150$ & $1.45 \pm 0.3$ & $1.90 \pm 0.3$ & $-$3.77 & 84 \\ 
Scl~6\_6\_402   & 01:00:00.39 & $-33$:29:15.2 & 19.13 & 1.07 & 2.19 & $4945 \pm 150$ & $2.00 \pm 0.3$ & $1.80 \pm 0.3$ & $-$3.53 & 76 \\ 
Scl~07-50       & 01:00:01.12 & $-33$:59:21.4 & 18.67 & 1.16 & 2.37 & $4558 \pm 150$ & $1.05 \pm 0.3$ & $2.35 \pm 0.3$ & $-$4.05 & 190 \\ 
Scl~07-49       & 01:00:05.00 & $-34$:01:16.6 & 18.38 & 1.19 & 2.46 & $4495 \pm 150$ & $0.80 \pm 0.3$ & $2.65 \pm 0.3$ & $-$3.31 & 204 \\ 
S1020549        & 01:00:47.83 & $-33$:41:03.2 & 18.34 & 1.20 & 2.51 & $4702 \pm 150$ & $1.25 \pm 0.3$ & $2.30 \pm 0.3$ & $-$3.68 & 171 \\
\enddata

\tablecomments{Coordinates and optical magnitudes are taken from
  \citet{coleman05}.  $K_{{\rm s}}$ magnitudes are from
  \citet{menzies11}.}
\tablenotetext{a}{The signal-to-noise ratio is given per \AA, measured
  at a wavelength of 5350~\AA.}
\label{sparams}
\end{deluxetable*}

\begin{deluxetable*}{llccccc}
\tablenum{4}
\tabletypesize{\scriptsize}
\tablewidth{0pt}
\tablecaption{Abundance Uncertainties (abridged)}
\label{uncert}
\tablehead{
\colhead{} &
\colhead{} &
\colhead{} &
\colhead{$\Delta$T$_{\rm eff}$} &
\colhead{$\Delta$log\,g} &
\colhead{$\Delta$v$_{\rm t}$} & 
\colhead{}\\
\colhead{Star} & \colhead{[X/Fe]} & \colhead{$\sigma_{{\rm obs}}$\tablenotemark{a}} &
\colhead{$+$150 K} &
\colhead{$+$0.3 dex} &
\colhead{$+$0.3 km s$^{-1}$} & \colhead{Total\tablenotemark{b}}}
\startdata
Scl 11\_1\_4296 & CH    & 0.20 & $+$0.26 & $-$0.09 & $-$0.01 & 0.34 \\
                & \ion{Na}{1} & 0.07 & $-$0.04 & $+$0.03 & $+$0.08 & 0.12 \\
                & \ion{Mg}{1} & 0.06 & $-$0.16 & $+$0.00 & $+$0.03 & 0.17 \\
                & \ion{Al}{1} & 0.20 & $-$0.04 & $+$0.02 & $+$0.08 & 0.22 \\
                & \ion{Si}{1} & 0.20 & $-$0.04 & $+$0.00 & $-$0.01 & 0.20 \\
                & \ion{Ca}{1} & 0.20 & $-$0.01 & $-$0.02 & $-$0.06 & 0.21 \\
                & \ion{Sc}{2} & 0.20 & $-$0.08 & $+$0.12 & $-$0.02 & 0.25 \\
                & \ion{Ti}{1}  & \nodata & \nodata & \nodata & \nodata & \nodata \\
                & \ion{Ti}{2} & 0.04 & $-$0.12 & $+$0.13 & $+$0.06 & 0.19 \\
                & \ion{Cr}{1} & 0.11 & $-$0.01 & $+$0.02 & $+$0.06 & 0.13 \\
                & \ion{Mn}{1} & 0.07 & $+$0.01 & $+$0.02 & $+$0.06 & 0.09 \\
                & \ion{Fe}{1}\tablenotemark{c} & 0.04 & $+$0.19 & $-$0.06 & $-$0.14 & 0.25\\
                & \ion{Fe}{2}\tablenotemark{c} & 0.07 & $+$0.03 & $+$0.09 & $-$0.03 & 0.12\\
                & \ion{Co}{1} & \nodata & \nodata & \nodata & \nodata & \nodata\\
                & \ion{Ni}{1} & \nodata & \nodata & \nodata & \nodata &\nodata \\
                & \ion{Zn}{1} & \nodata & \nodata & \nodata & \nodata & \nodata\\
                & \ion{Sr}{2} & \nodata & $-$0.09 & $+$0.11 & $+$0.09 & 0.17\\
                & \ion{Ba}{2} & \nodata & $-$0.01 & $+$0.16 & $+$0.19 & 0.25\\ 
                & \ion{Eu}{2} & \nodata & \nodata & \nodata & \nodata & \nodata\\
\hline
Scl 6\_6\_402   & CH    & 0.20 & $+$0.15 & $-$0.10 & $+$0.13 & 0.30 \\
                & \ion{Na}{1} & 0.11 & $-$0.03 & $+$0.02 & $+$0.04 & 0.12 \\
                & \ion{Mg}{1} & 0.09 & $+$0.01 & $-$0.08 & $+$0.00 & 0.12 \\
                & \ion{Al}{1} & 0.20 & $-$0.06 & $+$0.02 & $+$0.10 & 0.23 \\
                & \ion{Si}{1} & \nodata & $-$0.06 & $+$0.04 & $+$0.11 & 0.13 \\
                & \ion{Ca}{1} & 0.20 & $-$0.03 & $-$0.03 & $+$0.01 & 0.20 \\
                & \ion{Sc}{2} & 0.08 & $-$0.10 & $+$0.13 & $+$0.08 & 0.20 \\
                & \ion{Ti}{1}  & \nodata & \nodata & \nodata & \nodata & \nodata \\
                & \ion{Ti}{2} & 0.12 & $-$0.12 & $+$0.14 & $+$0.09 & 0.24 \\
                & \ion{Cr}{1} & \nodata & \nodata & \nodata & \nodata & \nodata \\
                & \ion{Mn}{1} & 0.13 & $-$0.01 & $+$0.02 & $+$0.09 & 0.16 \\
                & \ion{Fe}{1}\tablenotemark{c} & 0.04 & $+$0.21 & $-$0.04 & $-$0.13 & 0.25 \\
                & \ion{Fe}{2}\tablenotemark{c} & 0.05 & $+$0.03 & $+$0.10 & $-$0.04 & 0.12 \\
                & \ion{Co}{1} & 0.20 & $+$0.01 & $+$0.02 & $-$0.01 & 0.20 \\
                & \ion{Ni}{1} & 0.07 & $-$0.03 & $+$0.02 & $+$0.03 & 0.08 \\
                & \ion{Zn}{1} & \nodata & \nodata & \nodata & \nodata & \nodata \\
                & \ion{Sr}{2} & 0.20 & $-$0.09 & $+$0.20 & $+$0.03 & 0.30 \\
                & \ion{Ba}{2} & 0.20 & $-$0.08 & $+$0.10 & $+$0.06 & 0.24 \\
                & \ion{Eu}{2} & \nodata & $+$0.16 & $+$0.36 & $+$0.34 & 0.52 \\
\hline
Scl 07-50       & CH    & 0.20 & $+$0.22 & $-$0.11 & $+$0.08 & 0.34 \\
                & \ion{Na}{1} & 0.07 & $-$0.01 & $+$0.01 & $+$0.02 & 0.08 \\
                & \ion{Mg}{1} & 0.05 & $-$0.08 & $-$0.07 & $-$0.07 & 0.13 \\
                & \ion{Al}{1} & 0.07 & $-$0.03 & $-$0.03 & $+$0.00 & 0.08 \\
                & \ion{Si}{1} & 0.20 & $-$0.09 & $-$0.08 & $-$0.10 & 0.25 \\
                & \ion{Ca}{1} & 0.20 & $-$0.06 & $-$0.08 & $-$0.12 & 0.26 \\
                & \ion{Sc}{2} & 0.06 & $-$0.07 & $+$0.12 & $+$0.04 & 0.16 \\
                & \ion{Ti}{1}  & \nodata & \nodata & \nodata & \nodata & \nodata \\
                & \ion{Ti}{2} & 0.03 & $-$0.10 & $+$0.12 & $+$0.05 & 0.17 \\
                & \ion{Cr}{1} & 0.06 & $+$0.04 & $+$0.01 & $+$0.06 & 0.07 \\
                & \ion{Mn}{1} & 0.07 & $+$0.09 & $+$0.02 & $+$0.07 & 0.11 \\
                & \ion{Fe}{1}\tablenotemark{c} & 0.02 & $+$0.17 & $-$0.05 & $-$0.08 & 0.20 \\
                & \ion{Fe}{2}\tablenotemark{c} & 0.11 & $+$0.00 & $+$0.09 & $-$0.02 & 0.14 \\
                & \ion{Co}{1} & 0.06 & $+$0.06 & $+$0.01 & $+$0.05 & 0.08 \\
                & \ion{Ni}{1} & 0.06 & $+$0.03 & $-$0.01 & $+$0.00 & 0.07 \\
                & \ion{Zn}{1} & \nodata & \nodata & \nodata & \nodata & \nodata \\
                & \ion{Sr}{2} & 0.07 & $-$0.05 & $+$0.11 & $+$0.04 & 0.16 \\
                & \ion{Ba}{2} & 0.20 & $-$0.03 & $+$0.11 & $+$0.07 & 0.25 \\
                & \ion{Eu}{2} & \nodata & $-$0.06 & $+$0.11 & $+$0.08 & 0.38 \\
\enddata
\tablecomments{Only a portion of the table is shown in the arXiv version of
  the paper.  Please contact the first author or refer to the
  published version for the full table.}
\tablenotetext{a}{Dispersion of the abundance ratio about the mean for
  the species, divided by the square root of the number of lines,
  taken from Table~\ref{abundance_table}.  A minimum dispersion (before
  accounting for the number of lines) of 0.1~dex is imposed.  Species
  for which the abundance is determined from a single line are given a
  dispersion of 0.2~dex.}
\tablenotetext{b}{Sum in quadrature of $\sigma_{{\rm obs}}$,
  $\Delta \rm{T_{eff}}$, $\Delta \log{g}$, and $\Delta \rm{v_{t}}$.}
\tablenotetext{c}{[X/H] ratios are listed instead of [X/Fe] for
  \ion{Fe}{1} and \ion{Fe}{2}.}
\end{deluxetable*}

\subsection{Comparison to Literature Measurements}
\label{comparison}

Since three of the stars in our sample have been analyzed in the
literature using the same spectra we employ, we compare the abundance
results to check for differences resulting from assumptions or
methodology.  Our results for S1020549 agree with those of
\citet{fks10} within $1~\sigma$ (where $\sigma$ here refers to the
quadrature sum of the uncertainties in our measurements and those
determined by \citeauthor{fks10}) for all elements except Mg, where
our value of [Mg/Fe] is higher by 0.47~dex ($1.5~\sigma$).

Our agreement with T10 is reasonable, although slightly less good,
with 14 of 26 measurements (covering both Scl~07-50 and Scl~07-49)
matching to $1~\sigma$.  The \ion{Fe}{1} abundances we measure agree
with those determined by T10 ($\Delta \rm{[Fe/H]} = -0.09 \pm
0.21$~dex for Scl~07-50 and $\Delta \rm{[Fe/H]} = +0.17 \pm 0.20$~dex
for Scl~07-49), but with our stellar parameters there is no
significant difference between \ion{Fe}{1} and \ion{Fe}{2}.  Even
though these metallicity differences are not statistically
significant, they factor into more significant differences of
abundance ratios with respect to iron for some other elements (see
below).  We therefore conduct a detailed comparison of our [Fe/H]
measurements with those of T10 by examining the EWs of lines in
common.  Our EWs are systematically lower than those of T10, by an
average of $\sim3$~m\AA\ for all lines and $\sim5$~m\AA\ for Fe lines.
This EW offset would lead to a decrease of $\sim0.1-0.2$~dex in [Fe/H]
if the corresponding changes in stellar parameters were ignored.  We
note that our abundance determination uses many more lines than T10
did: 92 and 54 \ion{Fe}{1} lines for Scl~07-49 and Scl~07-50,
respectively, compared to 22 and 25.  We consider the most likely
explanation for the modest changes in [Fe/H] we derive to be a
combination of continuum placement, the lines used, and the different
methods for determining stellar parameters.

Of the differences in other elements for Scl~07-50 and Scl~07-49, most
are of modest significance ($\lesssim1.7~\sigma$).  The exceptions are
[Mg/Fe], where our measurements are higher by $1.7~\sigma$ and
$3.0~\sigma$, respectively, [Sc/Fe], where our measurements are lower
by $1.7-2.6~\sigma$, and [Ca/Fe] (for Scl~07-50 only), where our value
is $1.7~\sigma$ higher.  Addressing Mg first, the primary difference
for Scl~07-50 is our EW measurements, which are $\sim8$~m\AA\ larger.
This EW offset increases $\log{\epsilon\rm{(Mg)}}$ by 0.18~dex, which
when combined with our lower [Fe/H] results in [Mg/Fe] increasing by
0.27~dex.  Our $\log{gf}$ values for the Mg~b lines are smaller than
those used by T10, which accounts for the remainder of the increase in
[Mg/Fe].  For Scl~07-49 the offset in EWs relative to T10 is somewhat
smaller, but in the same sense.  The difference in stellar parameters
is responsible for changing the Mg abundance by 0.16~dex.  However,
our calculations of $\log{\epsilon\rm{(Mg)}}$ using T10's EWs and
stellar parameters yield $\log{\epsilon\rm{(Mg)}} = 4.62$, while their
reported value is 4.32, suggesting the possibility of unexpectedly
large differences between their model atmosphere calculations and
ours.

For Sc, the comparison is complicated by varying treatments of
hyperfine splitting and EW vs. spectral synthesis measurements, but
there are significant differences resulting from stellar parameters
(0.26~dex for Scl~07-50 and 0.13~dex for Scl~07-49).  The difference
in iron abundances makes the disagreement in [Sc/Fe] smaller than the
disagreement in $\log{\epsilon\rm{(Sc)}}$ for Scl~07-50, but the
reverse is true for Scl~07-49.  After factoring in changes in stellar
parameters and [Fe/H], our Sc abundances are still lower than those of
T10 by $0.2-0.3$~dex, which may result from differences in the
measured strength of the Sc lines and/or differences in stellar
atmospheres and analysis methods.  Finally, for Ca, our larger EW
results in an abundance that is higher by 0.21~dex.  This is partially
counteracted by the difference in stellar parameters, which lower
$\log{\epsilon\rm{(Ca)}}$, but as with Mg the combination of a higher
Ca abundance and a lower Fe abundance results in a substantial shift
in [Ca/Fe].  Also similar to Mg, we find that even if we use T10's EWs
and stellar parameters we get a value of $\log{\epsilon\rm{(Ca)}}$
that differs from theirs by 0.18~dex.

\section{DISCUSSION}
\label{discussion}

\subsection{Iron Peak Elements}

The newly discovered EMP stars Scl~6\_6\_402 and Scl~11\_1\_4296 have
${\rm [Fe/H]} = -3.53$ and ${\rm [Fe/H]} = -3.77$, respectively,
ranking both among the ten lowest metallicity stars known beyond the
Milky Way.

For the two stars previously studied by T10, as mentioned in
\S~\ref{comparison}, we find \ion{Fe}{1} abundances that are in
agreement within the uncertainties.  Our measurements are ${\rm
  [Fe/H]} = -3.31$ for Scl~07-49 and ${\rm [Fe/H]} = -4.05$ for
Scl~07-50.  We note that while Scl~07-50 is only slightly more
metal-poor in our analysis, it now formally qualifies as the first
ultra metal-poor star identified in an external galaxy according to
the nomenclature proposed by \citet{bc05}.  For S1020549 we obtain
${\rm [Fe/H]} = -3.68$, in agreement with our earlier determination
within the uncertainties \citep{fks10}.

\begin{figure*}[t!]
\epsscale{1.24}
\plotone{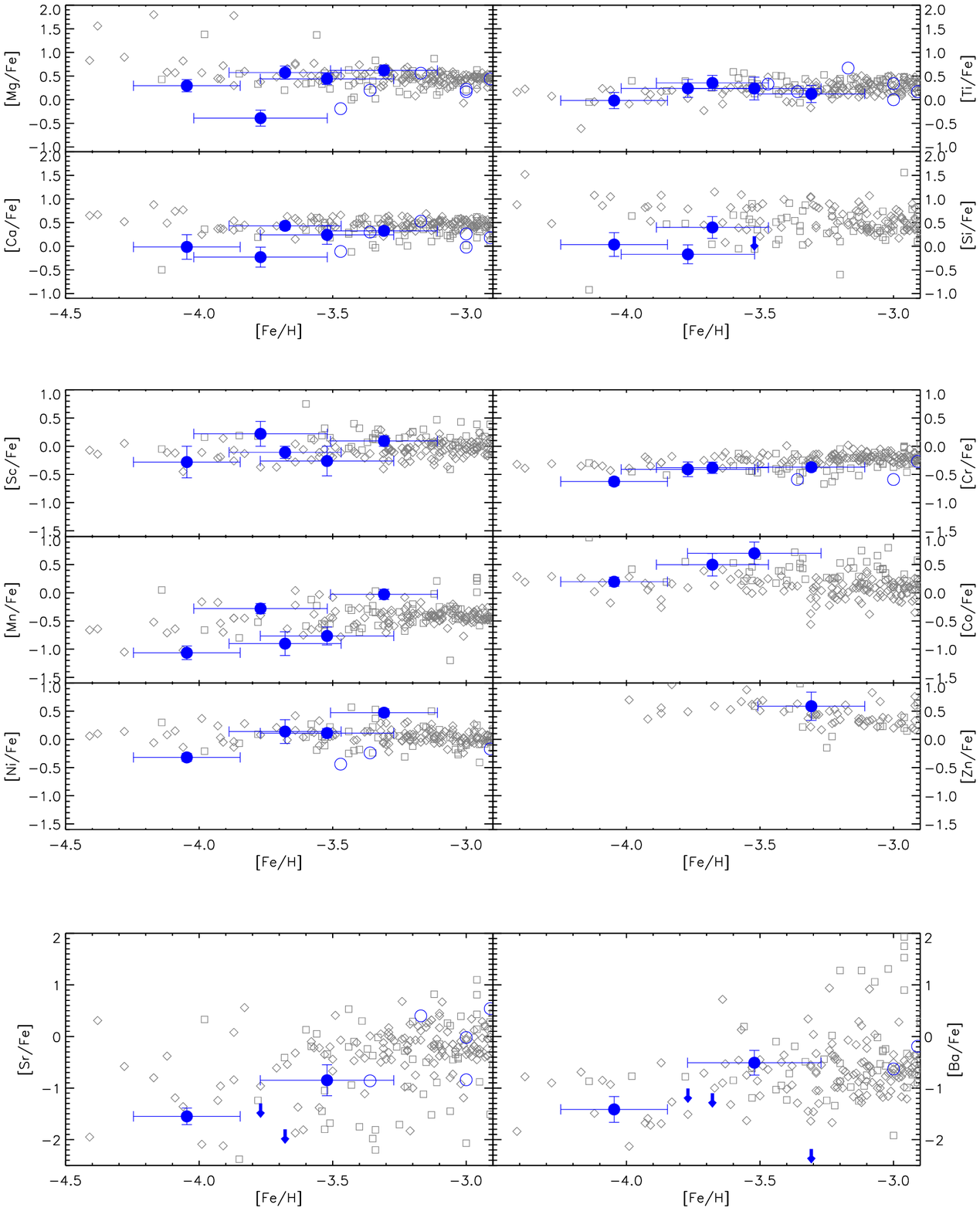}
\caption{Abundance patterns of EMP stars in Sculptor (filled blue
  circles) compared to the Milky Way halo samples of \citet[][open
    gray squares]{cohen13} and \citet[][open gray
    diamonds]{roederer14}.  The Sculptor sample of
  \citet{starkenburg13} is plotted as open blue circles.  The
  $\alpha$-elements are displayed in the top set of panels, iron peak
  elements in the middle set of panels, and neutron-capture elements
  in the bottom panels.  The \citet{cohen13} abundances have been
  adjusted to place them on the \citet{asplund09} solar abundance
  scale.}
\label{abundances}
\end{figure*}

For the other iron-peak elements Sc, Cr, Mn, Co, Ni, and Zn the
abundance patterns of the Scl stars are broadly consistent with those
seen in the Milky Way halo at similar metallicities (see
Figure~\ref{abundances}).  The Co abundances tend to be higher than
those of most halo stars, while the Mn abundances for three of the
Sculptor stars are near the lower bound of [Mn/Fe] observed in EMP
halo stars.  Scl~11\_1\_4296 and especially Scl~07-49 have Mn
abundances above the halo average, but the value for the latter star
is determined only from two weak lines because of the lack of blue
spectral coverage.  Our spectra of the other four stars extend
to shorter wavelengths, allowing us to use the much stronger
\ion{Mn}{1}~$\lambda$4030~\AA\ triplet resonance lines.  The
corrections for non-LTE (NLTE) behavior in these lines are significant
\citep[e.g.,][]{cayrel04,lai08,bg08}, and we adopt an NLTE correction
of $+0.30$~dex to abundances determined from the Mn~triplet.  The only
other potential outlier among the iron-peak species is the low Ni
abundance of Scl~07-50, which is comparable to the lowest [Ni/Fe]
values seen at ${\rm [Fe/H]} < -3.5$ in the halo.

\subsection{$\alpha$-Elements}
\label{alphas}

The abundance patterns of the $\alpha$-elements in the Sculptor EMP
stars closely follow the well-established behavior of EMP Milky Way
halo stars
\citep[e.g.,][]{mcwilliam95,cayrel04,cohen13,yong13,roederer14} down
to $\rm{[Fe/H]} = -3.7$.  Above this metallicity Mg, Ca, Si, and
Ti lie almost perfectly along the median of the Milky Way halo
distribution (see Figure~\ref{abundances}).  However, the two lowest
metallicity stars show some striking differences, both from each other
and from their more metal-rich counterparts: Scl~11\_1\_4296 has
uniformly low $\alpha$ abundances (except Ti), while Scl~07-50 has low
Ca and Si but an almost normal [Mg/Fe] ratio.

\begin{figure*}[t!]
\epsscale{1.24}
\plotone{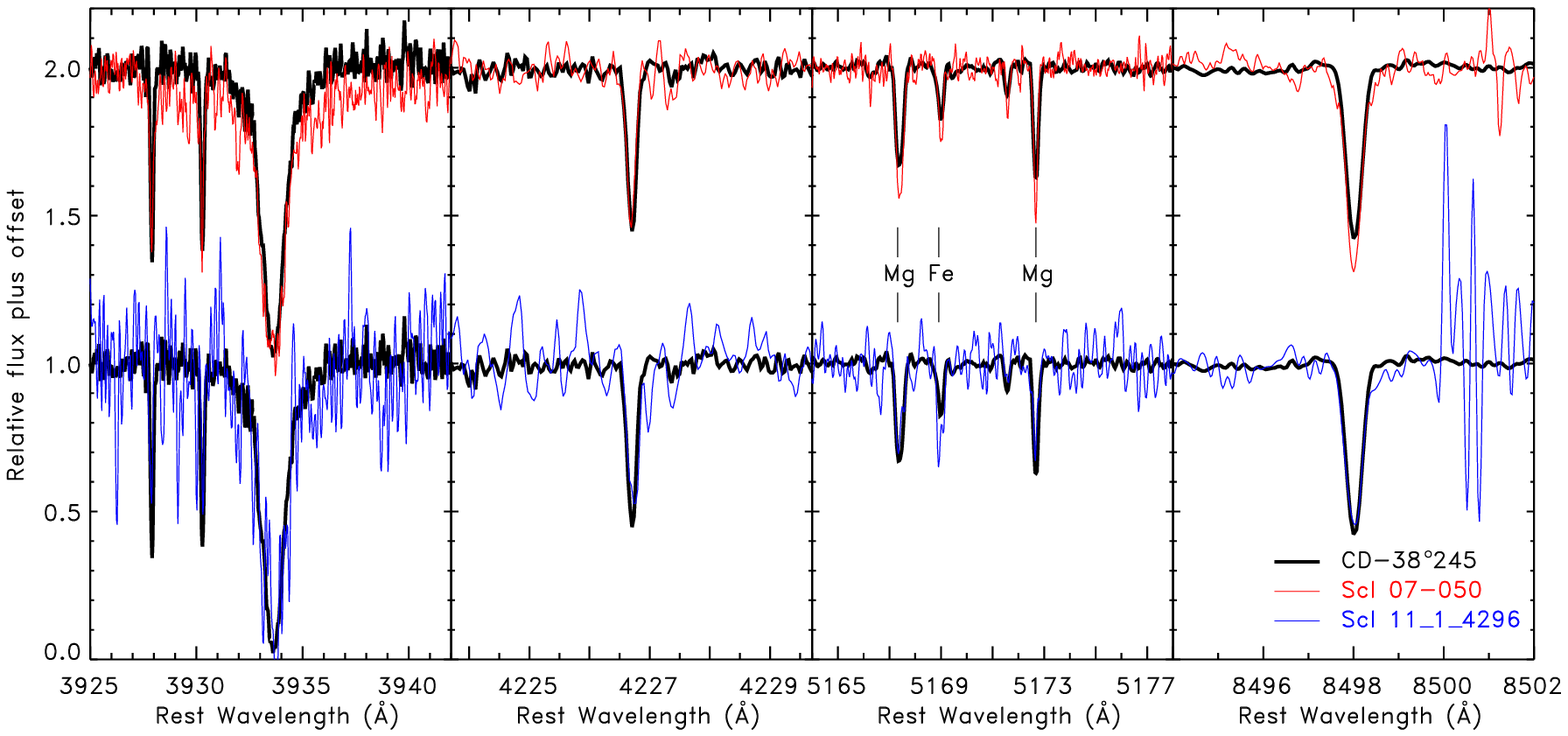}
\caption{Spectra of the two most metal-poor stars in Sculptor,
  Scl~07-50 (red) and Scl~11\_1\_4296 (blue, smoothed slightly for
  cosmetic purposes), compared to CD$-38\degr$245.  Ca~K is plotted in
  the left panel, \ion{Ca}{1}~4226.73~\AA\ in the second panel, the
  bluer two lines of the Mg triplet in the third panel, and the bluest
  \ion{Ca}{2} triplet line in the right panel.}
\label{spectra}
\end{figure*}

For the stars below $\rm{[Fe/H]} = -3.7$, our Ca abundances are
determined from the only \ion{Ca}{1} line detected, the
4226.73~\AA\ resonance line.  This line is known to produce lower Ca
abundances than other \ion{Ca}{1} lines for EMP stars, at least in
part because of NLTE effects \citep[e.g.,][]{spite12}, leading T10 to
dismiss the significance of the even lower [Ca/Fe] ratio they derived
for Scl~07-50.  However, NLTE models do not agree well on the
correction for the 4226.73~\AA\ line for stars with similar
atmospheric parameters to Scl~07-50 and Scl~11\_1\_4296, with recent
predictions ranging from $-0.02$~dex \citep{starkenburg10} to
$+0.21$~dex \citep*[][L. Mashonkina 2014, personal
  communication]{mashonkina07}.  We therefore attempted several
additional tests to verify the low Ca abundances.  First, we compared
the Ca lines of both stars with those of the ultra metal-poor giant
CD$-38\degr$245 \citep{bn84}, which is comparable in temperature to
Scl~11\_1\_4296 and $\sim200$~K warmer than Scl~07-50 (see
Figure~\ref{spectra}).  Scl~11\_1\_4296 has weaker
\ion{Ca}{1}~$\lambda$4226.73~\AA\ and near-infrared \ion{Ca}{2}
triplet lines than CD$-38\degr$245, confirming its low Ca abundance.
Scl~07-50 has similar Ca~K and \ion{Ca}{1}~$\lambda$4226.73~\AA\ EWs
to CD$-38\degr$245, consistent with a lower Ca abundance given the
temperature difference.\footnote{The Ca triplet lines of Scl~07-50 are
  stronger than those of CD$-38\degr$245, but NLTE and 3D corrections
  for those lines are much larger and even less well understood.}
Second, we compared to stars with similar parameters (after adjusting
their spectroscopic temperatures according to the \citealt{frebel13}
formula) from \citet{roederer14}.  These stars were selected to have
4400~K~$ < \rm{T_{eff}} <$~4900~K, $-0.2 < \log{g} < 1.8$,
$\rm{[Fe/H]} < -3$, and a detection of the \ion{Ca}{1} resonance line,
resulting in a sample of 13 stars.  For this sample, Ca abundances
from \ion{Ca}{1}~4226.73~\AA\ are 0.09~dex lower than the mean
abundance from all other \ion{Ca}{1} lines.\footnote{The only outlier
  where \ion{Ca}{1}~4226.73~\AA\ and the other lines have a
  significantly larger abundance difference is the coolest star,
  CS~22950-046, which is similar in temperature to Scl~07-50, raising
  the possibility of a sharp temperature dependence in the abundance
  derived from the resonance line.  However, because the theoretical
  studies of \citet{mashonkina07}, \citet{merle11}, and
  \citet{spite12} do not indicate strong changes in NLTE corrections
  at $\rm{T_{eff}} \sim 4500$~K, we regard CS~22950-046 as a random
  outlier rather than a systematic one.}  This offset is in excellent
agreement with the most recent NLTE corrections for Ca determined by
\citet{spite12}, which give 0.08~dex for T$_{{\rm eff}} = 4750$~K,
$\log{g} = 1$, $\rm{[Ca/H]} = -3.2$.  Finally, we stacked the four
strongest \ion{Ca}{1} non-resonance lines and determined upper limits
of $\rm{[Ca/Fe]} < -0.06$~dex and $\rm{[Ca/Fe]} < -0.12$~dex for
Scl~07-50 and Scl~11\_1\_4296, respectively, by comparing to
synthesized spectra.  We therefore conclude that even after factoring
in the uncertain NLTE effects, Scl~07-50 and Scl~11\_1\_4296 indeed
have low Ca abundances.

Using the large halo samples of \citet{cohen13} and \citet{roederer14}
as a guide, only one other star with such a low Ca abundance at ${\rm
  [Fe/H]} < -3.5$ is known: HE~1424-0241, which has a similar
metallicity but an even more extreme Ca underabundance of ${\rm
  [Ca/Fe]} = -0.50$ \citep{cohen07,cohen13}.  The other 40 halo stars
in this metallicity range are all at ${\rm [Ca/Fe]} \ge 0.15$.
HE~1424-0241 is not a perfect analog for the Sculptor stars because it
is even more under-abundant in Si than in Ca, at ${\rm [Si/Fe]} =
-1.01$, and has normal Mg and enhanced Mn and Co.  Scl~11\_1\_4296, on
the other hand, is deficient in Mg ($\rm{[Mg/Fe]} = -0.39$) as well as
the other $\alpha$-elements, while Scl~07-50 has a low Si abundance
but is not nearly as depleted as HE~1424-0241, and its Mn abundance is
on the low end compared to typical halo stars (see above).
\citet{cohen07} argue that HE~1424-0241 must have been enriched by a
very small number of supernovae (SNe), with the Si-deficient material
likely contributed by a supernova of around 35~M$_{\odot}$.

The other known stars with $\alpha$-element abundance patterns
reminiscent of Scl~07-50 are the two stars in the ultra-faint dwarf
galaxy Hercules studied by \citet{koch08}.  Those stars are
substantially more metal-rich, at $\rm{[Fe/H]} \approx -2$, but have
high Mg and O abundances combined with low Ca, making for extreme
[Mg/Ca] ratios.  As with HE~1424-0241, \citet{koch08} suggest that
this abundance pattern results from small numbers of SN explosions and
a high-mass ($\sim35$~M$_{\odot}$) progenitor.

\subsection{Carbon}

Neither of the two new Sculptor stars presented in this paper is
carbon-rich (for Scl~6\_6\_402 we obtain only an upper limit on
[C/Fe], but that is sufficient to rule out a substantial carbon
enhancement).  \citet{starkenburg13} recently analyzed a sample of
seven Sculptor stars with $-3.5 < \rm{[Fe/H]} < -2.5$, plus S1020549
and Scl~07-50, similarly finding no examples of stars with large
carbon enhancements.  They concluded that there is potential tension
between the lack of identified carbon-enhanced metal-poor (CEMP) stars
in Sculptor and the prevalence of such stars in the Milky Way, but a
larger sample is needed for this result to be conclusive (also see
\citealt{skuladottir14}).  Since we only add two additional
non-carbon-enhanced EMP stars, the probability calculations they
reported do not change significantly.  A sample roughly twice as large
will be necessary to determine whether the fraction of CEMP stars in
Sculptor is truly discrepant with that of the halo.

\subsection{Neutron-Capture Elements}

Only a few neutron-capture species can be measured in our spectra.  We
detect the \ion{Sr}{2} resonance lines in Scl~6\_6\_402 and Scl~07-50,
with low abundances of $\rm{[Sr/Fe]} = -0.85$ and $\rm{[Sr/Fe]} = 
  -1.55$ ($\rm{[Sr/H]} = -4.37$ and $\rm{[Sr/H]} = -5.60$),
respectively.  S1020549 has an even lower upper limit of $\rm{[Sr/Fe]}
< -1.66$ ($\rm{[Sr/H]} < -5.48$).  Ba is detected in the same two
stars, with [Ba/Fe] ratios and upper limits that are mostly closer to
the halo averages than those for Sr.  Eu upper limits rule out large
enhancements of r-process elements but are not otherwise constraining.

These results generally fit in with the picture of very low abundances
of neutron-capture elements in the most metal-poor stars in dwarf
galaxies \citep{fulbright04,fks10,fsk14,ishigaki14}.  Sr in particular
seems notably underabundant in our Sculptor sample.

\subsection{Implications for the Early History of Sculptor}

Considering our sample in conjunction with the seven additional
Sculptor stars from \citet{starkenburg13}, there is evidence for
abundance spreads within the Sculptor EMP population in
$\alpha$-elements (Mg and Ca), iron peak elements (Ni), and
neutron-capture elements (Sr and Ba).  While comparisons to the Milky
Way halo are hampered by the smaller sample size in Sculptor, and more
importantly the lower data quality, Figure~\ref{abundances} suggests
that the star-to-star scatter among Sculptor stars for a number of
elements may be higher than in the halo.  To investigate the
significance of the increased scatter, we calculated the intrinsic
scatter in [X/Fe] for both EMP stars in the halo data sets of
\citet{cohen13} and \citet{roederer14} and the combined Sculptor
sample from \citet{starkenburg13} and this paper.  We used the method
described by \citet{kelly07} to determine the intrinsic scatter in a
set of data points with non-uniform uncertainties in both variables.
The elements for which halo EMP stars follow tight relations and there
are more than five measurements available for Sculptor stars are Mg,
Ca, Ti, Cr, and Ni.

For Mg, the handful of stars with highly enhanced abundances at
$\rm{[Fe/H]} < -3.5$ give the halo sample a significant intrinsic
scatter of 0.17~dex.  Still, the very low [Mg/Fe] ratio of
Scl~11\_1\_4296 and the low Mg abundances of several of the
\citet{starkenburg13} stars result in a larger scatter for Sculptor of
$\sigma = 0.36^{+0.14}_{-0.10}$~dex, which is significant at the 99\%
confidence level.  The scatter of the halo stars in [Ca/Fe] is
slightly smaller (0.15~dex), but the outliers in Sculptor are less
deviant, making the intrinsic scatter of the Sculptor stars larger at
only 71\% confidence.  [Ti/Fe] has even less scatter in the halo data
(observed $\sigma_{\rm{[Ti/Fe]}} = 0.14$~dex, intrinsic
$\sigma_{\rm{[Ti/Fe]}} = 0.10$~dex).  The Sculptor stars generally
follow the halo trend in [Ti/Fe] closely, but the intrinsic scatter of
$0.24^{+0.14}_{-0.11}$~dex is larger at 91\% confidence.  For Cr, we
similarly find that the intrinsic scatter of the Sculptor stars of
$\sigma = 0.18^{+0.16}_{-0.10}$ is larger than that of the halo
($\sigma = 0.10$~dex) at 80\% confidence.  Finally, for the iron-peak
element Ni the larger uncertainties in [Ni/Fe] leave very little room
for any intrinsic scatter in the halo sample; we derive an intrinsic
scatter of $0.04 \pm 0.02$~dex.  The Sculptor stars, in contrast, have
$\sigma = 0.46^{+0.35}_{-0.19}$, which is larger at 99.9\% confidence.
Given the still small sample of metal-poor stars in Sculptor, these
differences are (with the exception of Mg and Ni) suggestive rather
than definitive, and abundance patterns for more Sculptor stars will
be needed to confirm their greater diversity relative to the halo.
Nevertheless, the apparently increased heterogeneity in the most
metal-poor Sculptor stars highlights the likely role played by
inhomogeneous mixing in the early Sculptor interstellar medium and the
small number of SNe that contributed to its initial chemical
enrichment.

The abundance patterns of the two most metal-poor stars also point to
enrichment by a limited number of progenitor stars.  The low Si, Ca,
and Ni abundances of Scl~07-50 and the exceptionally low $\alpha$
abundances of Scl~11\_1\_4296 stand out strongly from the typical
abundances of similar metallicity stars in the halo, which presumably
represent the mean yield from a larger number of primordial SNe.  To
determine what kind of SNe could be responsible for this chemical
makeup, we fit their abundances with the Population~III SN models of
\citet{hw10}.  The best fitting models for Scl~07-50 are all
relatively high mass (22.5~M$_{\odot}$) stars, while Scl~11\_1\_4296
is better fit by a $\sim10$~M$_{\odot}$ star.  Models with lower mass
progenitors for Scl~07-50 and higher mass progenitors for
Scl~11\_1\_4296 have significantly higher $\chi^{2}$ values.  In both
cases, the data are consistent with high energy hypernova explosions
with $E \gtrsim 3 \times 10^{51}$~erg, although lower energy
explosions are not excluded for Scl~11\_1\_4296.  It is intriguing
that significantly different progenitor properties are preferred for
these two stars.  Since the SNe from a normal stellar initial mass
function (IMF) will be dominated by the lowest mass stars that can
undergo core collapse ($\sim8-10$~M$_{\odot}$), the larger mass for
the star that enriched Scl~07-50 may suggest that the IMF in Sculptor
at the earliest times was biased toward high masses.  \citet{geha13}
have shown that the ultra-faint dwarfs Hercules and Leo~IV have a
bottom-light IMF, which if extrapolated to the high mass regime would
be top-heavy.  No direct measurements for Sculptor have been made yet,
but would clearly be of interest.

To quantify how many supernovae could have contributed to the
enrichment of the $\alpha$-poor EMP stars in Sculptor, we followed the
method described by \citet{koch08}.  We randomly generated small
populations of massive stars ($10~{\rm M_{\odot}} \le M \le 100~{\rm
  M_{\odot}}$) and used the \citet{hw10} metal-free supernova yields
(with a randomly chosen explosion energy and mixing parameter for each
star) to determine the mean abundance ratios that would result after
these stars polluted a primordial gas cloud.  For a \citet{salpeter}
IMF, sub-solar [Ca/Fe] ratios like we observe for Scl~11\_1\_4296 are
found in less than 1\% of these simulations unless the number of
supernovae is less than five.  ${\rm [Ca/Fe]} < -0.2$ is exceedingly
rare even for a single supernova explosion.  More top-heavy IMFs,
which may be appropriate for dwarf galaxies
\citep{wyse02,kalirai13,geha13}, modestly reduce the likelihood of
producing extremely Ca-poor material.  We conclude that Sculptor may
have hosted only $1-4$ SNe at the time that its most metal-poor known
stars were formed.

\section{SUMMARY AND CONCLUSIONS}
\label{conclusions}

We have determined chemical abundances for five of the most metal-poor
stars in the Sculptor dSph, all at $\rm{[Fe/H]} < -3.2$.  Two of these
stars are new discoveries reported here for the first time, while
three others are taken from the literature and re-analyzed.  Our
slightly revised metallicity for one of the literature stars is below
$\rm{[Fe/H]} = -4$, classifying it as the first ultra metal-poor star
known in a galaxy other than the Milky Way.

The two lowest metallicity stars both have very unusual abundance
patterns.  The $\alpha$ elements Mg, Ca, and Si are all extremely
depleted in Scl~11\_1\_4296, while Scl~07-50 has low Ca, Si, and Ni,
but a more normal level of Mg.  The unique abundance patterns of
the most metal-poor stars suggests that $\rm{[Fe/H]} \lesssim -3.7$ is
the regime in which only a few SNe contributed to the enrichment of
Sculptor, implying that the stars studied here are part of just the
second generation the galaxy ever formed.  These chemical signatures
can be reproduced by enrichment from Population~III SNe.

The Ca abundances of the two most metal-poor stars rely entirely on
the \ion{Ca}{1} resonance line at 4226\AA, which is the only neutral
Ca line detected in our spectra.  Because the effects of NLTE on this
line are uncertain, we compared the \ion{Ca}{2} K, \ion{Ca}{1} 4226,
and \ion{Ca}{2} triplet lines in the Sculptor stars with the ultra
metal-poor giant CD$-38\degr$245, finding that the abundances we
derive are qualitatively consistent with the relative line strengths.
We then used a set of EMP halo stars with similar metallicity and
temperature in which multiple \ion{Ca}{1} lines are detected to show
that the resonance line produces a Ca abundance that is on average
0.09~dex lower than the abundance determined from other lines.  This
value is consistent with the latest NLTE calculations.

Although the three more metal-rich stars individually appear quite
similar to Milky Way halo stars in the same metallicity range, when
taken together with other Sculptor EMP stars we show that the
star-to-star scatter in the abundance of several elements is
substantially higher than in the halo.  We infer that the early
chemical evolution of Sculptor was heavily influenced by inhomogeneous
mixing and stochastic effects from small numbers of SNe.

\acknowledgements{This publication is based upon work supported by the
  National Science Foundation under grants AST-1108811 and
  AST-1255160.  We thank the referee for suggestions that strengthened
  the paper.  We thank Atish Kamble for contributing observations,
  Gary da Costa for the photometric catalog, and Andy McWilliam,
  Lyudmila Mashonkina, Juna Kollmeier, and Ian Roederer for helpful
  conversations.  This research has made use of NASA's Astrophysics
  Data System Bibliographic Services.  }

{\it Facilities:} \facility{Magellan:II (MIKE)}

\clearpage
\pagestyle{empty}
\begin{turnpage}
\begin{deluxetable}{llcccccccccccc}
\tablecolumns{14}
\tablewidth{0pt}
\tabletypesize{\small}
\tablecaption{Equivalent Widths (abridged)}
\tablehead{
\colhead{Species}  &
\colhead{$\lambda$}  &
\colhead{$\chi$}  &
\colhead{$\log{gf}$}  &
\colhead{EW (m\AA)}  &
\colhead{$\log{\epsilon}$ (dex) } &
\colhead{EW (m\AA)}  &
\colhead{$\log{\epsilon}$ (dex) } &
\colhead{EW (m\AA)}  &
\colhead{$\log{\epsilon}$ (dex) } &
\colhead{EW (m\AA)}  &
\colhead{$\log{\epsilon}$ (dex) } &
\colhead{EW (m\AA)}  &
\colhead{$\log{\epsilon}$ (dex) } \\
\multicolumn{1}{c}{} &
\multicolumn{1}{c}{[\AA]} &
\multicolumn{1}{c}{[eV]} &
\multicolumn{1}{c}{[dex]} &
  \multicolumn{2}{c}{Scl~11\_1\_4296} &
  \multicolumn{2}{c}{Scl~6\_6\_402} &
  \multicolumn{2}{c}{Scl~07-50} &
  \multicolumn{2}{c}{Scl~07-49} &
  \multicolumn{2}{c}{S1020549} }
\startdata
         CH & 4313.00 & \nodata &  \nodata & \phn syn &             5.00 & \phn syn &          $<$5.50 & \phn syn &    \phm{$<$}4.10 &  \nodata &          \nodata & \phn syn &          $<$4.95 \\
         CH & 4325.00 & \nodata &  \nodata &  \nodata &          \nodata & \phn syn &          $<$5.50 & \phn syn &          $<$4.47 &  \nodata &          \nodata & \phn syn &          $<$5.25 \\
\ion{Na}{1} & 5889.95 &    0.00 & \phs0.11 & \phn79.3 &             2.40 & \phn97.5 &    \phm{$<$}2.90 & \phn82.3 &    \phm{$<$}2.19 &    123.3 &    \phm{$<$}2.72 &    102.4 &    \phm{$<$}2.68 \\
\ion{Na}{1} & 5895.92 &    0.00 &  $-$0.19 & \phn65.3 &             2.45 & \phn71.2 &    \phm{$<$}2.69 & \phn59.1 &    \phm{$<$}2.12 &    106.6 &    \phm{$<$}2.73 & \phn83.1 &    \phm{$<$}2.64 \\
\ion{Mg}{1} & 3829.36 &    2.71 &  $-$0.21 & \phn83.4 &             3.61 &  \nodata &          \nodata & \phn98.7 &    \phm{$<$}3.80 &  \nodata &          \nodata &  \nodata &          \nodata \\
\ion{Mg}{1} & 3832.30 &    2.71 & \phs0.27 & \phn97.0 &             3.46 &  \nodata &          \nodata &  \nodata &          \nodata &  \nodata &          \nodata &  \nodata &          \nodata \\
\ion{Mg}{1} & 3838.29 &    2.72 & \phs0.49 &  \nodata &          \nodata &  \nodata &          \nodata &    131.9 &    \phm{$<$}3.86 &  \nodata &          \nodata &  \nodata &          \nodata \\
\ion{Mg}{1} & 5172.68 &    2.71 &  $-$0.45 & \phn65.9 &             3.32 &    119.9 &    \phm{$<$}4.43 &    107.4 &    \phm{$<$}3.87 &    185.4 &    \phm{$<$}4.95 &    129.3 &    \phm{$<$}4.39 \\
\ion{Mg}{1} & 5183.60 &    2.72 &  $-$0.24 & \phn78.7 &             3.36 &    145.5 &    \phm{$<$}4.61 &    117.6 &    \phm{$<$}3.87 &    196.0 &    \phm{$<$}4.87 &    154.3 &    \phm{$<$}4.60 \\
\ion{Mg}{1} & 5528.40 &    4.34 &  $-$0.50 &  \nodata &          \nodata &  \nodata &          \nodata &  \nodata &          \nodata & \phn55.8 &    \phm{$<$}4.92 &  \nodata &          \nodata \\
\ion{Al}{1} & 3944.01 &    0.00 &  $-$0.62 &  \nodata &          \nodata & \phn syn &    \phm{$<$}2.02 & \phn syn &    \phm{$<$}1.32 &  \nodata &          \nodata & \phn syn &          $<$2.00 \\
\ion{Al}{1} & 3961.52 &    0.01 &  $-$0.34 & \phn59.4 &             1.68 & \phn syn &          $<$2.42 & \phn syn &    \phm{$<$}1.40 &  \nodata &          \nodata & \phn syn &          $<$1.90 \\
\ion{Si}{1} & 3905.52 &    1.91 &  $-$1.09 & \phn87.2 &             3.57 & \phn syn &          $<$4.20 & \phn syn &    \phm{$<$}3.50 &  \nodata &          \nodata & \phn syn &    \phm{$<$}4.23 \\
\ion{Si}{1} & 4102.94 &    1.91 &  $-$3.14 &  \nodata &          \nodata & \phn syn &          $<$5.08 & \phn syn &          $<$3.92 &  \nodata &          \nodata & \phn syn &          $<$4.50 \\
\ion{Ca}{1} & 4226.73 &    0.00 & \phs0.24 & \phn97.5 &             2.25 &    126.4 &    \phm{$<$}2.97 &    108.6 &    \phm{$<$}2.19 &  \nodata &          \nodata &    149.2 &    \phm{$<$}3.11 \\
\ion{Ca}{1} & 4434.96 &    1.89 &  $-$0.01 &  \nodata &          \nodata & \phn57.3 &          $<$4.32 &  \nodata &          \nodata &  \nodata &          \nodata & \phn32.4 &    \phm{$<$}3.15 \\
\ion{Ca}{1} & 5588.76 &    2.52 & \phs0.21 &  \nodata &          \nodata &  \nodata &          \nodata &  \nodata &          \nodata & \phn24.8 &    \phm{$<$}3.35 &  \nodata &          \nodata \\
\ion{Ca}{1} & 5594.47 &    2.52 & \phs0.10 &  \nodata &          \nodata &  \nodata &          \nodata &  \nodata &          \nodata &  \nodata &          \nodata & \phn13.1 &    \phm{$<$}3.23 \\
\ion{Ca}{1} & 6102.72 &    1.88 &  $-$0.79 &  \nodata &          \nodata &  \nodata &          \nodata &  \nodata &          \nodata & \phn19.1 &    \phm{$<$}3.40 &  \nodata &          \nodata \\
\ion{Ca}{1} & 6122.22 &    1.89 &  $-$0.32 &  \nodata &          \nodata &  \nodata &          \nodata &  \nodata &          \nodata & \phn37.4 &    \phm{$<$}3.32 & \phn16.0 &    \phm{$<$}2.99 \\
\ion{Ca}{1} & 6162.17 &    1.90 &  $-$0.09 &  \nodata &          \nodata &  \nodata &          \nodata &  \nodata &          \nodata & \phn57.3 &    \phm{$<$}3.42 &  \nodata &          \nodata \\
\ion{Ca}{1} & 6439.07 &    2.52 & \phs0.47 &  \nodata &          \nodata &  \nodata &          \nodata &  \nodata &          \nodata & \phn37.2 &    \phm{$<$}3.30 & \phn17.7 &    \phm{$<$}2.99 \\
\ion{Sc}{2} & 4246.82 &    0.32 & \phs0.24 & \phn81.6 &          $-$0.40 & \phn syn & \phm{$<$}$-$0.80 & \phn syn & \phm{$<$}$-$1.13 &  \nodata &          \nodata & \phn syn & \phm{$<$}$-$0.73 \\
\ion{Sc}{2} & 4314.08 &    0.62 &  $-$0.10 &  \nodata &          \nodata & \phn syn & \phm{$<$}$-$0.55 & \phn syn & \phm{$<$}$-$1.20 &  \nodata &          \nodata & \phn syn & \phm{$<$}$-$0.63 \\
\ion{Sc}{2} & 4325.00 &    0.59 &  $-$0.44 &  \nodata &          \nodata & \phn syn & \phm{$<$}$-$0.55 & \phn syn & \phm{$<$}$-$1.20 &  \nodata &          \nodata & \phn syn & \phm{$<$}$-$0.73 \\
\ion{Sc}{2} & 4400.39 &    0.61 &  $-$0.54 &  \nodata &          \nodata &  \nodata &          \nodata &  \nodata &          \nodata &  \nodata &          \nodata & \phn syn & \phm{$<$}$-$0.78 \\
\ion{Sc}{2} & 4415.54 &    0.59 &  $-$0.67 &  \nodata &          \nodata &  \nodata &          \nodata & \phn syn &       $<$$-$0.74 &  \nodata &          \nodata & \phn syn & \phm{$<$}$-$0.31 \\
\ion{Sc}{2} & 5031.01 &    1.36 &  $-$0.40 &  \nodata &          \nodata &  \nodata &          \nodata & \phn syn &       $<$$-$0.44 &      syn & \phm{$<$}$-$0.18 &  \nodata &          \nodata \\
\ion{Sc}{2} & 5526.79 &    1.77 & \phs0.02 &  \nodata &          \nodata &  \nodata &          \nodata &  \nodata &          \nodata &      syn &    \phm{$<$}0.05 &  \nodata &          \nodata \\
\ion{Sc}{2} & 5657.91 &    1.51 &  $-$0.60 &  \nodata &          \nodata &  \nodata &          \nodata &  \nodata &          \nodata &      syn &          $<$0.11 &  \nodata &          \nodata \\
\enddata
\label{ew_table}
\tablecomments{Only a portion of the table is shown in the arXiv version of
  the paper.  Please contact the first author or refer to the
  published version for the full table.}
\end{deluxetable}
\end{turnpage}
\clearpage
\global\pdfpageattr\expandafter{\the\pdfpageattr/Rotate 90}

\clearpage
\pagestyle{empty}
\begin{turnpage}
\begin{deluxetable}{l | rrcc | rrcc | rrcc | rrcc  | rrcc }
\tablenum{3}
\tablewidth{0pt}
\setlength{\tabcolsep}{.1cm}
\tablecaption{Abundances}
\label{abundance_table}
\tablehead{
  \multicolumn{1}{c}{Species} &
  \multicolumn{4}{c}{Scl~11\_1\_4296} &
  \multicolumn{4}{c}{Scl~6\_6\_402} &
  \multicolumn{4}{c}{Scl~07-50} &
  \multicolumn{4}{c}{Scl~07-49} &
  \multicolumn{4}{c}{S1020549} \\
\colhead{} & 
\colhead{[X/Fe]} & \colhead{log$\epsilon(X)$} & \colhead{No.} & \colhead{$\sigma$\tablenotemark{a}} &  
\colhead{[X/Fe]} & \colhead{log$\epsilon(X)$} & \colhead{No.} & \colhead{$\sigma$\tablenotemark{a}} &         
\colhead{[X/Fe]} & \colhead{log$\epsilon(X)$} & \colhead{No.} & \colhead{$\sigma$\tablenotemark{a}} &
\colhead{[X/Fe]} & \colhead{log$\epsilon(X)$} & \colhead{No.} & \colhead{$\sigma$\tablenotemark{a}} &
\colhead{[X/Fe]} & \colhead{log$\epsilon(X)$} & \colhead{No.} & \colhead{$\sigma$\tablenotemark{a}} \\
\colhead{} & \colhead{[dex]} & \colhead{[dex]} & \colhead{lines} & \colhead{[dex]} &
\colhead{[dex]} & \colhead{[dex]} & \colhead{lines} & \colhead{[dex]} &
\colhead{[dex]} & \colhead{[dex]} & \colhead{lines} & \colhead{[dex]} &
\colhead{[dex]} & \colhead{[dex]} & \colhead{lines} & \colhead{[dex]} &
\colhead{[dex]} & \colhead{[dex]} & \colhead{lines} & \colhead{[dex]} 
}
\startdata 
C(CH)       & \phs0.34 & 5.00  & 1  & \nodata             & $<0.59$  & $<5.50$    &  1 & \nodata   & $-0.28$  & 4.10 &  1 & \nodata         & \nodata  & \nodata & \nodata & \nodata & $<0.20$  & $<4.95$ &  1 & \nodata  \\
\ion{Na}{1} & $-0.04$  & 2.43  & 2  & 0.03                & \phs0.08 & 2.80    &  2 & 0.15         & $-0.03$  & 2.16 &  2 & 0.05            & $-0.20$  & 2.73 &  2 & 0.01            & \phs0.10 & 2.66 &  2 & 0.03  \\  
\ion{Mg}{1} & $-0.39$  & 3.44  & 4  & 0.11                & \phs0.44 & 4.52    &  2 & 0.13         & \phs0.30 & 3.85 &  4 & 0.03            & \phs0.62 & 4.91 &  3 & 0.04            & \phs0.58 & 4.50 &  2 & 0.15  \\  
\ion{Al}{1} & $-1.00$  & 1.68  & 1  & \nodata             & $-0.91$  & 2.02    &  1 & \nodata      & $-1.04$  & 1.36 &  2 & 0.06            & \nodata  & \nodata & \nodata & \nodata & $<-0.87$ & $<1.90$ &  1 & \nodata  \\
\ion{Si}{1} & $-0.17$  & 3.57  & 1  & \nodata             & $<0.21$  & $<4.20$    &  1 & \nodata   & \phs0.04 & 3.50 &  1 & \nodata         & \nodata  & \nodata & \nodata & \nodata & \phs0.40 & 4.23 &  1 & \nodata  \\
\ion{Ca}{1}\tablenotemark{b} & $-0.23$  & 2.34  & 1  & \nodata             & \phs0.24 & 3.06    &  1 & \nodata      & $-0.01$ & 2.28 &  1 & \nodata         & \phs0.33 & 3.36 &  5 & 0.05            & \phs0.43 & 3.09 &  5 & 0.10  \\   
\ion{Sc}{2} & \phs0.22 & $-0.40$ & 1 & \nodata            & $-0.26$  & $-0.63$ &  3 & 0.14         & $-0.28$  & $-1.18$ & 3 & 0.04          & \phs0.09 & $-0.07$ &  2 & 0.16         & $-0.11$  & $-0.64$ & 5 & 0.19  \\  
\ion{Ti}{1} & \nodata  & \nodata & \nodata & \nodata      & \nodata  & \nodata & \nodata & \nodata & \nodata  & \nodata & \nodata & \nodata & \phs0.06 & 1.70 &  5 & 0.09            & \phs0.48 & 1.75 &  2 & 0.01  \\   
\ion{Ti}{2} & \phs0.23 & 1.41  & 11 & 0.12                & \phs0.24 & 1.67    &  4 & 0.23         & $-0.01$ & 0.89 & 10 & 0.11            & \phs0.19 & 1.83 &  6 & 0.10            & \phs0.34 & 1.61 & 18 & 0.22  \\    
\ion{Cr}{1} & $-0.41$  & 1.46  & 3  & 0.19                & \nodata  & \nodata & \nodata & \nodata & $-0.62$  & 0.97 &  4 & 0.12            & $-0.37$  & 1.96 &  2 & 0.06            & $-0.38$  & 1.58 &  5 & 0.20  \\    
\ion{Mn}{1}\tablenotemark{c} & $-0.28$ & 1.38 & 2  & 0.10 & $-0.77$  & 1.14    &  3 & 0.23         & $-1.06$  & 0.32 &  3 & 0.16            & $-0.02$  & 2.10 &  2 & 0.11            & $-0.90$  & 0.85 &  1 & \nodata  \\    
\ion{Fe}{1}\tablenotemark{d} & $-3.77$ & 3.73 & 54 & 0.32 & $-3.52$  & 3.98    & 44 & 0.24         & $-4.05$  & 3.45 & 54 & 0.10            & $-3.31$  & 4.19 & 92 & 0.14            & $-3.68$  & 3.82 & 84 & 0.24  \\  
\ion{Fe}{2}\tablenotemark{d} & $-3.76$ & 3.74 & 2  & 0.10 & $-3.52$  & 3.98    &  4 & 0.09         & $-4.04$  & 3.46 &  4 & 0.13            & $-3.28$  & 4.22 &  5 & 0.11            & $-3.67$  & 3.83 &  5 & 0.16  \\   
\ion{Co}{1} & \nodata  & \nodata & \nodata & \nodata      & \phs0.70 & 2.17    &  1 & \nodata      & \phs0.20 & 1.14 &  3 & 0.08            & \nodata  & \nodata & \nodata & \nodata & \phs0.50 & 1.81 &  1 & \nodata  \\
\ion{Ni}{1} & \nodata  & \nodata & \nodata & \nodata      & \phs0.11 & 2.81    &  2 & 0.10         & $-0.32$  & 1.85 &  3 & 0.11            & \phs0.48 & 3.39 &  4 & 0.14            & \phs0.14 & 2.68 &  1 & \nodata  \\   
\ion{Zn}{1} & \nodata  & \nodata & \nodata & \nodata      & \nodata & \nodata & \nodata & \nodata  & \nodata  & \nodata & \nodata & \nodata & \phs0.59 & 1.84 &  1 & \nodata         & \nodata & \nodata & \nodata & \nodata \\   
\ion{Sr}{2} & $<-1.30$ & $<-2.20$ & 1 & \nodata           & $-0.85$  & $-1.50$ &  1 & \nodata      & $-1.55$  & $-2.73$ & 2 & 0.04          & \nodata  & \nodata & \nodata & \nodata & $<-1.80$ & $<-2.61$ &  1 & \nodata  \\
\ion{Ba}{2} & $<-1.01$ & $<-2.60$ & 1 & \nodata           & $-0.51$  & $-1.85$ &  1 & \nodata      & $-1.41$  & $-3.28$ & 1 & \nodata      & $<-2.18$ & $<-3.31$ &  1 & \nodata     & $<-1.10$ & $<-2.60$ &  1 & \nodata  \\   
\ion{Eu}{2} & \nodata  & \nodata & \nodata & \nodata      & $<1.29$  & $<-1.71$ & 1 & \nodata      & $<0.76$  & $<-2.77$ & 1 & \nodata      & \nodata  & \nodata & \nodata & \nodata & $<1.00$  & $<-2.16$ &  1 & \nodata  \\
\enddata
\tablenotetext{a}{Dispersion of the abundance ratio about the mean for
  species where multiple lines were detected.}
\tablenotetext{b}{An NLTE correction of $+0.09$~dex was applied to
  \ion{Ca}{1} abundances determined only from the 4226.73~\AA\ line
  (see \S~\ref{alphas}).}
\tablenotetext{c}{An NLTE correction of $+0.3$~dex was applied to
  abundances determined from the 4030~\AA\ triplet lines.}
\tablenotetext{d}{Abundances are relative to hydrogen for \ion{Fe}{1}
  and \ion{Fe}{2}.}
\end{deluxetable}
\end{turnpage}
\clearpage
\global\pdfpageattr\expandafter{\the\pdfpageattr/Rotate 90}

\end{document}